\documentclass[conference]{IEEEtran}

\usepackage[utf8]{inputenc}
\usepackage{todonotes}
\usepackage{enumitem}
\usepackage{booktabs}
\usepackage{tabularx}
\usepackage{multirow}
\usepackage{url}

\usepackage{mdframed}
\mdfsetup{leftmargin=3pt,linewidth=1pt}
\usepackage{float}

\let\oldsim\sim 
\renewcommand{\sim}{{\oldsim}}

\title{Sustainability in Software Architecture:\\ A Systematic Mapping Study}

\author{\IEEEauthorblockN{Vasilios Andrikopoulos}
    \IEEEauthorblockA{\textit{University of Groningen} \\
    Groningen, the Netherlands \\
    v.andrikopoulos@rug.nl}
    \and
    \IEEEauthorblockN{Rares-Dorian Boza}
    \IEEEauthorblockA{\textit{University of Groningen} \\
    Groningen, the Netherlands \\
    r.d.boza@student.rug.nl}
    \and
    \IEEEauthorblockN{Carlos Perales}
    \IEEEauthorblockA{\textit{Vrije Universiteit Amsterdam} \\
    Amsterdam, the Netherlands \\
    c.peraleslinan@student.vu.nl}
    \and
    \IEEEauthorblockN{Patricia Lago}
    \IEEEauthorblockA{\textit{Vrije Universiteit Amsterdam} \\
    Amsterdam, the Netherlands \\
    p.lago@vu.nl}
}

\begin{document}
\bstctlcite{IEEEexample:BSTcontrol} 

\maketitle

\begin{abstract}
Sustainability is an increasingly-studied topic in software engineering in general, and in software architecture in particular.
%
There are already a number of secondary studies addressing sustainability in software engineering, but no such study focusing explicitly on software architecture. 
This work aims to fill this gap by conducting a systematic mapping study on the intersection between sustainability and software architecture research with the intention of (i) reflecting on the current state of the art, and (ii) identifying the needs for further research.
Our results show that, overall, existing works have focused disproportionately on specific aspects of sustainability, and in particular on the most technical and ``inward facing'' ones. 
This comes at the expense of the holistic perspective required to address a multi-faceted concern such as sustainability.
Furthermore, more reflection-oriented research works, and better coverage of the activities in the architecting life cycle are required to further the maturity of the area.
Based on our findings we then propose a research agenda for sustainability-aware software architecture.
\end{abstract}

\section{Introduction}
\label{sec:intro}

The developments of the last decades have generated a raising awareness of sustainability in an increasingly digital world run by software systems~\cite{Becker2014,GeSIwPurpose}.
With software systems being deeply embedded in most sectors of our society, a unique opportunity is being offered for a shared intervention across sectors towards a more sustainable world~\cite{Penzenstadler2014}. 
%
Reflecting this need, in the recent years sustainability has been acknowledged as an essential software quality across four dimensions~\cite{Lago2015}:
a \emph{technical} one referring to the ability of a software system to evolve and remain used over a long period of time,
an \emph{economic} one concerned with the preservation and creation of capital and value,
an \emph{environmental} one aiming to minimize the impact of the system on natural resources,
and a \emph{social} one focusing on the continuity of communities using it.

The state of the art in sustainable software engineering in general has evolved significantly in the last couple of decades, starting with a focus on `green' software, that is, focusing on the environmental dimension, e.g.~\cite{Naumann2011,Mahmoud2013,Kern2015,Hindle2016}, 
but expanding also to the other ones, e.g.~\cite{penzenstadler2012,Garcia-Mireles2018, Wolfram2018}.
A recent survey of sustainable software research based on the 5Ws formula (why, when, who, where, and what)~\cite{Calero2020}, shows that this is a very active and collaborative area of research, with a good level of maturity.
The research community, or more precisely its intersection with that of software engineering, is also mapped out in terms of major outlets and collaborative efforts by the surveys of Calero and Piattini~\cite{Calero2017} and Lago and Penzenstadler~\cite{Lago2017}, indicating the existence of active research groups in the area.

Despite however the availability of secondary studies such as the ones mentioned above discussing sustainability in software engineering, there are fewer works describing the state of the art in the intersection between sustainability and software architecture.
Focusing specifically on architecture is essential since architecting practices allow us to reason and evaluate sustainability as a system quality throughout the system's lifecycle.
This is especially important since there is still a lot of ground to be covered with respect to establishing and exploiting the relation between software architecture and sustainability as a software quality~\cite{Venters2018}.

The secondary studies that do exist on the topic have their own limitations.
The survey by Venters et al.~\cite{Venters2018}, for example, provides a good overview of this area but it is not conducted systematically.
Some surveys either focus on technical sustainability, e.g.~the one by Koziolek on sustainability evaluation metrics for software architectures~\cite{koziolek2011sustainability}, or the ones on the sustainability of reference architectures~\cite{volpato2017two} and their description~\cite{valle2021towards}. 
Verdecchia et al.~\cite{verdecchia2018architectural} discuss technical in combination with economic sustainability in the context of architectural technical debt, 
while Volpato et al.~\cite{Volpato2019} and Grua et al.~\cite{Grua2021} look into the implications of software architecture on the social dimension. 
%
No systematic survey so far, to the extent of our knowledge, attempts to \emph{organize the state of the art of sustainability in software architecture across all sustainability dimensions and architecting activities}.
This is a need that this work aims to address by mapping out the research efforts in the field and thus identifying areas of future research.
For this purpose, in this paper we report on our \emph{systematic mapping study} on this topic, following well-established guidelines~\cite{Petersen2008systematic,petersen2015guidelines} for the study design, execution, and reporting.

\begin{figure*}
    \centering
    \includegraphics[width=.85\textwidth]{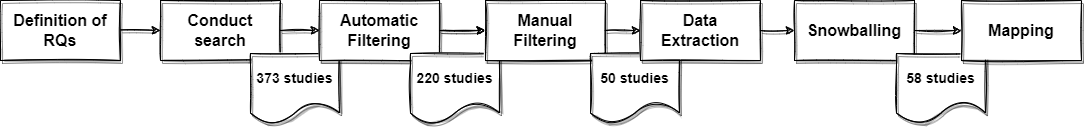}
    \caption{The study protocol as a flowchart.}
    \label{fig:protocol}
\end{figure*}

As a consequence, the rest of this paper is organized as follows.
Section~\ref{sec:design} discusses the methodological aspects of this study and presents the actions we took as part of conducting it.
Section~\ref{sec:findings} summarizes the findings of the study, aiming to answer the research questions we have identified in the process.
Section~\ref{sec:discussion} synthesizes these findings into a research agenda for sustainability-aware software architecture for the coming years.
Finally, Section~\ref{sec:validity} discusses the threats to the validity of this work, and Section~\ref{sec:conclusion} concludes it with a summary of its main points.

\section{Study Design}
\label{sec:design}

\subsection{Methodology}

To conduct our mapping study in a systematic manner, we are following the guidelines of Petersen et al.~\cite{Petersen2008systematic,petersen2015guidelines} for this type of studies in software engineering.
The systematic mapping process prescribed by these guidelines can be roughly broken down into the following phases:
\begin{enumerate}
    \item Definition of research questions to scope the study.
    \item Searching and filtering of relevant papers through well defined inclusion and exclusion criteria.
    \item Classification of the selected papers through keywording of the abstracts.
    \item Extraction of relevant data and preparation of the map.
\end{enumerate}
For the purposes of our study we have adopted this process with some amendments as shown in Fig.~\ref{fig:protocol}.
More specifically, keyword extraction from abstracts alone proved insufficient in the majority of cases to properly characterize the primary studies, pushing their characterization to the next step (data extraction). 
The same insufficiency also applied in many cases to the filtering of papers, requiring the reading of the whole paper to control for some of the inclusion/exclusion criteria we defined.
We therefore found more natural to finalize the filtering of the papers together with doing data extraction.
This last stage of filtering was performed manually and complements a previous stage of automated filtering for criteria that could be checked through scripting.
Furthermore, and as a result of doing filtering and extraction together, we also had to push snowballing~\cite{jalali2012systematic} for locating additional relevant studies at a later stage,  doing data extraction separately for those studies.

In the following we discuss each of these phases in detail.

\subsection{Definition of Research Questions}

As the goal of this work is to provide an overview of the research on sustainability in software architecture we pose questions from three distinct perspectives.
First, we ask what are the overall publication trends in this research area in the last years, and how mature is the related literature.
Second, from a software architecture perspective, we are interested in which activities of the architecting life cycle~\cite{hofmeister2007,tang2010comparative} are taken into consideration when sustainability is of concern for the systems under study.
Third, from a sustainability research perspective, we want to know which of the dimensions of sustainability as a software quality~\cite{Lago2015} are discussed in the literature.

These perspectives translate into the following research questions:
\begin{mdframed}
	\begin{enumerate}[label=\textbf{RQ\textsubscript{\arabic*}},ref=RQ\textsubscript{\arabic*}]
		\item What are the existing approaches related to sustainability in the field of software architecture as reflected by the literature?\label{rq:overall}
		\item To what extent are the activities of the architecting life cycle covered by these approaches?\label{rq:phases}
		\item To what extent are the dimensions of sustainability covered by the same approaches?\label{rq:dims}
	\end{enumerate}
\end{mdframed}

\subsection{Search Process}

As per the recommendation of Petersen et al.~\cite{petersen2015guidelines}, we focus only on the Population and Intervention dimensions of the PICO strategy~\cite{kitchenham2007guidelines} for identifying keywords and formulating search strings from the research questions.
More specifically, as the intended Population of our study we define \textit{``all primary studies discussing the architecture of sustainable software systems and relevant activities''} where an Intervention of \textit{``one or more dimensions of sustainability as a software quality is the target of the presented approach''} takes place.
Based on this definition and after extensive piloting to calibrate the efficacy of our search we decided on using the query:
\begin{quote}
    \texttt{\footnotesize "software architect*" AND sustainab* AND (requirements OR design OR implementation OR evaluation OR decision)}
\end{quote}
which allows searching for publications using variations of the main concepts (sustainable and sustainability, and software architecture, architecting, and architects) in combination with activities that are indicative of actually discussing software architecture (e.g. eliciting requirements, system design, etc.)
Five online bibliographical sources were used for the search as shown in Table~\ref{tab:search}. 
\begin{table}[b]
	\vspace{-10pt}
	\centering
	\caption{Search results for the defined query per online bibliographical source}
	\label{tab:search}
	\begin{tabular}{lc}
		\toprule
		\bfseries Source & \bfseries \#Publications \\
		\midrule
		ACM Digital Library & 22\\
		IEEE eXplore & 138\\
		Science Direct & 20\\
		Scopus & 187\\
		Wiley Online Library & 6\\
		\midrule
		\bfseries Total: & 373\\
		\bottomrule
	\end{tabular}
\end{table}
The query was adjusted accordingly for each source based on the syntax and limitations of their respective search engines.
The search took place in June 2021. 
It was additionally constrained to publications after 2000 since previous secondary studies on sustainability in software engineering~\cite{penzenstadler2012,Wolfram2018,Calero2020} report findings only after this year anyway.
Table~\ref{tab:search} shows the result of this search per source, for a total of 373 candidate publications identified by our search.

\subsection{Publications Filtering}

In alignment with our research questions we defined the following inclusion criteria for this study:
\begin{enumerate}[label=\textbf{IC\textsubscript{\arabic*}},ref=IC\textsubscript{\arabic*},left=0.5em]
    \item Publication focuses on software architecture.\label{ic:architecture}
    \item Publication is related to one or more sustainability dimensions.\label{ic:sustainability}
    \item Publication is peer reviewed (journal articles, conference and workshop proceedings, magazine articles).\label{ic:peer}
\end{enumerate}

A publication was excluded from our study if one or more of the following criteria were satisfied:
\begin{enumerate}[label=\textbf{EC\textsubscript{\arabic*}},ref=EC\textsubscript{\arabic*},left=0.5em]
    \item Publication is not published in English.
    \item The publication full text is not publicly accessible.
    \item Publication is a duplicate of another study.
    \item Publication is a secondary study.\label{ec:secondary}
\end{enumerate}

Both the last inclusion and the first three exclusion criteria can be checked quite easily automatically.
For this purpose we created a set of scripts that took the search result lists from the previous step as an input, retrieved all references and publication metadata in a unified format, merged the lists and removed duplicates, and checked the language and type of the publication based on its bibliographical data.
Publications that could not be automatically retrieved were flagged for possible manual retrieval, with 71 publications being found to be completely irretrievable for various reasons.
This resulted in a set of 220 publications after this automated filtering for further processing as shown in Fig.~\ref{fig:protocol}.

The last exclusion criterion (\ref{ec:secondary}), however, and the first two inclusion criteria (\ref{ic:architecture} and~\ref{ic:sustainability}) required reading the whole publication in order to decide if it was admissible to the study.
Especially the latter two criteria proved quite elusive and needed further deliberations for their application.
Ultimately, we consider that a publication focuses on software architecture \emph{if we are able to identify which architectural activities it concerns itself with}. 
This is irrespective of whether the work in question is explicitly contributing to the study of software architectures in general, or discusses a reusable architecting approach to a specific problem across a whole domain.
The relation to sustainability, on the other hand, is established through the primary study providing sufficient details on \emph{how its content relates to one or more of the sustainability dimensions as they pertain to software systems}.
With respect to the latter point we found quite common for publications to use sustainability as a keyword in their opening and/or closing sections without however providing any connection to the concept throughout their main body. 
As a result we ended up removing quite a lot of publications from the final sample due to their superficial relation with sustainability.
Out of the 220 publications from the automated filtering step, 50 were selected for inclusion in our study through this process.

Since both the identification of the discussed sustainability dimension(s) and the architecting phase(s) is part of the data extraction, as will be discussed further in the following, we ended up doing this second stage of filtering together with the data extraction one.
In order to minimize the introduced bias, two co-authors worked on these two tasks independently but at the same time. 
Disagreements between them concerning the inclusion/exclusion of a primary study were resolved in iterative consensus meetings mediated by a third author who also read through all publications.

\subsection{Data Extraction \& Snowballing}

\begin{table}
    \centering
    \caption{Data extraction form used in this study; fields marked with asterisk extracted automatically}
    \label{tab:form}
    \renewcommand{\arraystretch}{1.2}
    \begin{tabularx}{\columnwidth}{lXc}
        \toprule
        \bfseries Field & \bfseries Values & \bfseries RQ \\
        \midrule
         Publication Venue* & Full name of venue & \multirow{8}{*}{\ref{rq:overall}}\\
         Venue Type* & Journal or Conference & \\
         Publication Date* & Year of publication & \\  
         Keywords & Comma-separated list of keywords & \\
         \multirow{3}{*}{Research Facet} & One of the facets identified by Wieringa et al.~\cite{wieringa2006requirements}, as per~\cite{Petersen2008systematic} & \\
         \multirow{2}{*}{Contribution Viewpoint} & One of the viewpoints identified in Venters et al.~\cite{Venters2018} & \\
         \midrule
         \multirow{3}{*}{Architecting Phases} & One or more from the phases identified by Tang et al.~\cite{tang2010comparative} & \multirow{2}{*}{\ref{rq:phases}}\\
         \midrule
         \multirow{3}{*}{Sustainability Dimensions} & One or more from the dimensions identified by Lago et al.~\cite{Lago2015} & \multirow{2}{*}{\ref{rq:dims}}\\ 
         \bottomrule
    \end{tabularx}
\vspace{-10pt}
\end{table}
Table~\ref{tab:form} summarizes the form used for data extraction from the (primary) studies, following the example of Petersen et al.~\cite{Petersen2008systematic}.
Keywords were extracted from the title and abstract of each study, reusing in some, but not all cases the keywords provided in the publication itself.
Obvious keywords such as `software architecture' or `sustainability' and derivatives were not included since they were used for the filtering of the study.
Publication venue types were bundled into two categories: \emph{journals}, including also magazines, and \emph{conferences}, including also workshop proceedings.
Together with the publication date, these venue-related fields were extracted automatically from the metadata of the publications with occasional manual interventions for providing missing data.
Research facets use the classification of Wieringa et al.~\cite{wieringa2006requirements} to categorize research efforts into \emph{validation or evaluation research}, \emph{solution proposals}, or \emph{philosophical, opinion, or experience papers}.
Following Venters et al.~\cite{Venters2018} we characterize primary studies depending on whether they discuss \emph{sustainable software}, where the software system and its architecting process is the goal of the study, or \emph{sustainability through software}, where delivering sustainability to the stakeholders is the product of this process.

Instead of the more widely accepted but less comprehensive categorization scheme of Hofmeister et al.~\cite{hofmeister2007}, we use the one by Tang et al.~\cite{tang2010comparative} to identify architecture-related activities in five phases: \emph{analysis}, \emph{synthesis} (design), \emph{evaluation}, \emph{implementation}, and \emph{maintenance} (covering also evolution).
To characterize the sustainability dimension(s) addressed by the primary study we use the Lago et al.~\cite{Lago2015} proposal of looking at the economic, environmental, social, and technical aspects of sustainability as a software quality. 
More than one phases or dimensions could be addressed by the same study, so both these fields are structured as comma-separated lists of elements from the respective categorization scheme. 

As discussed above, extracting data for these latter two fields provided us with the means of verifying~\ref{ic:architecture} and~\ref{ic:sustainability}: if no phase or dimension could be extracted for a study then it was treated as a strong indication that it had failed the respective inclusion criterion.
This however meant that we could perform a search for additional studies through backward snowballing~\cite{jalali2012systematic} only after the combined filtering and extraction was finalized.
In practice, this turned out not being a major issue; we found 8 additional studies by snowballing through the 50 studies that were filtered by this stage and carried out an additional extraction for them.
The results of the mapping process presented in the following section aggregate both sets of 58 primary studies.
The final list of primary studies, together with the outcomes of the previous steps of the process are available online\footnote{\url{https://figshare.com/s/a8a25157bf3feace9714}}.

\section{Findings}
\label{sec:findings}
\begin{figure}
	\centering
	\includegraphics[width=\columnwidth]{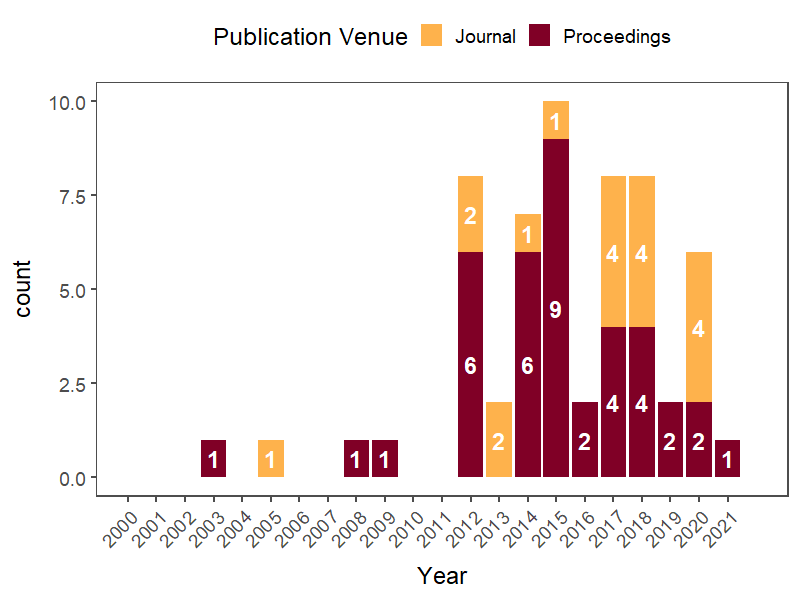}
	\caption{Publication year and venue type; proceedings include both conference and workshop proceedings.}
	\label{fig:pubs}
	\vspace{-10pt}
\end{figure}
\begin{table}
	\centering
	\caption{Publication venues with more than one study; conferences include also co-located workshops, except if published independently from the main conference.}
	\label{tab:top_venues}
	\renewcommand{\arraystretch}{1.2}
	\begin{tabularx}{\columnwidth}{Xcc}
		\toprule
		\bfseries Venue & \bfseries Type & \bfseries \#Studies \\
		\midrule
		European Conference on Software Architecture (ECSA) & \multirow{2}{*}{Conference} & \multirow{2}{*}{8}\\
		Working IEEE/IFIP Conference on Software Architecture (WICSA) --- later renamed to:  & \multirow{4}{*}{Conference} & \multirow{4}{*}{$7 (5+2)$}\\
		IEEE International Conference on Software Architecture (ICSA) & & \\
		IEEE Software & Journal & 4\\
		Information and Software Technology & Journal & 3\\
		IEEE Access & Journal & 2\\
		International Workshop on Requirements Engineering for Sustainable Systems (RE4SuSy) & \multirow{2}{*}{Conference} & \multirow{2}{*}{2}\\
		\bottomrule
	\end{tabularx}
	\vspace{-10pt}
\end{table}
%



\subsection{\ref{rq:overall} --- State of the Art overview}
\label{sec:rq1}
\begin{figure}
	\centering
	\includegraphics[width=\columnwidth]{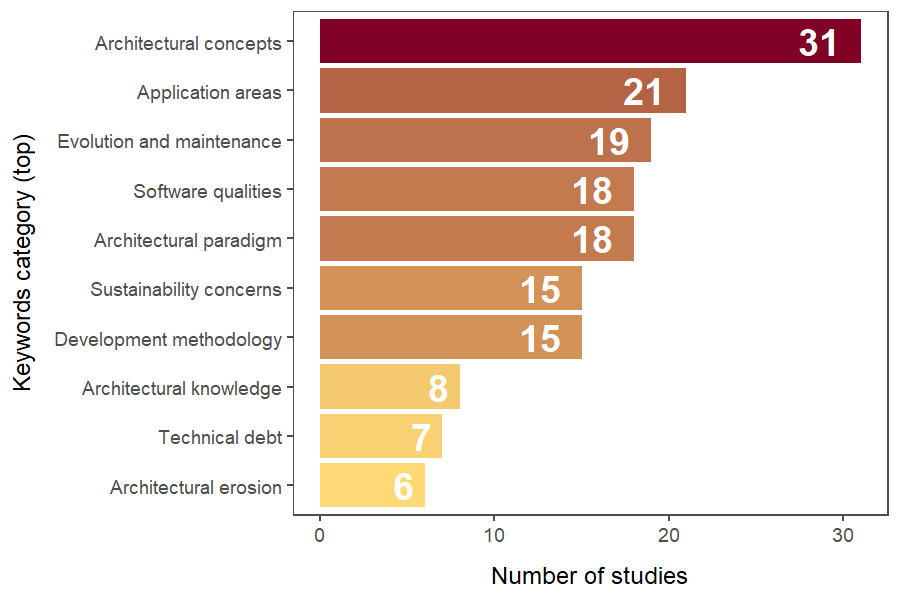}
	\caption{Top keywords categories after consolidation; multiple categories per study are possible.}
	\label{fig:keywords}
	\vspace{-10pt}
\end{figure}
\begin{figure}
	\centering
	\includegraphics[width=\columnwidth]{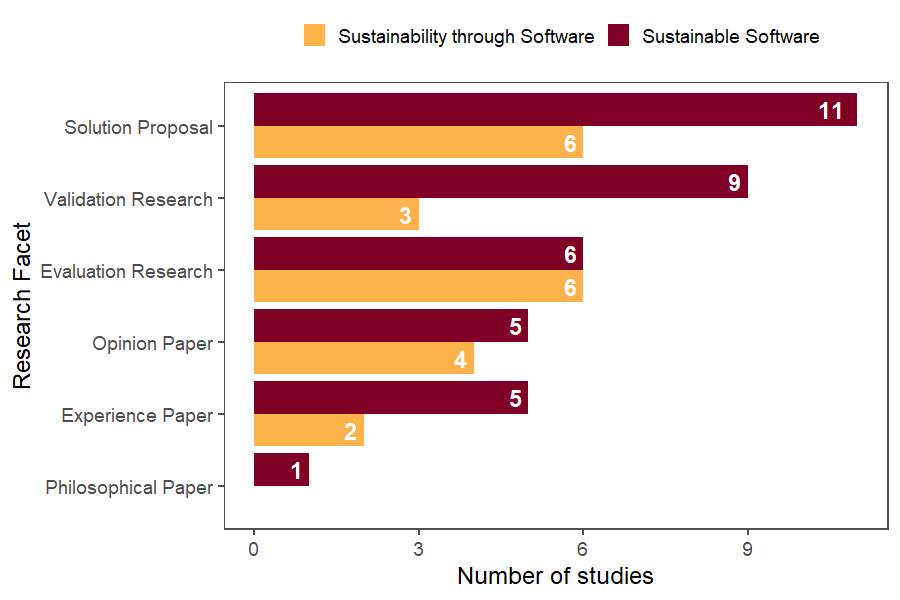}
	\caption{Research facets of the studies, together with their viewpoint with respect to sustainability}
	\label{fig:facets}
	\vspace{-10pt}
\end{figure}
To answer our first research question we start by looking at the publication of the selected studies using descriptive statistics. 
Fig.~\ref{fig:pubs} summarizes the number of publications per year after 2000 and per publication venue type: journal or magazine articles, and workshop or conference proceedings.
Publications in proceedings dominate with $67\%$ of the total, indicating space for more mature works to appear in the literature.
Table~\ref{tab:top_venues} outlines the publication venues with $n>1$ from the selected studies.
Perhaps unsurprisingly, the most popular venues for publishing on the topic appear to be the European and International Conferences on Software Architecture (ECSA and ICSA respectively), the latter in both its incarnations as WICSA and later ICSA.
The Software and Access IEEE journals, and the Information and Software Technology (IST) journal seem to be also preferred venues for publishing on the topic.
Of special mention is the International Workshop on Requirements Engineering for Sustainable Systems (Re4SuSy) run in conjunction with the Requirements Engineering (RE) conference.
The fact that the remaining 32 studies are published in 32 different venues is possibly a sign of a need for more concentrated targeting of the research output by the community. 

\begin{figure*}
	\centering
	\includegraphics[width=.7\textwidth]{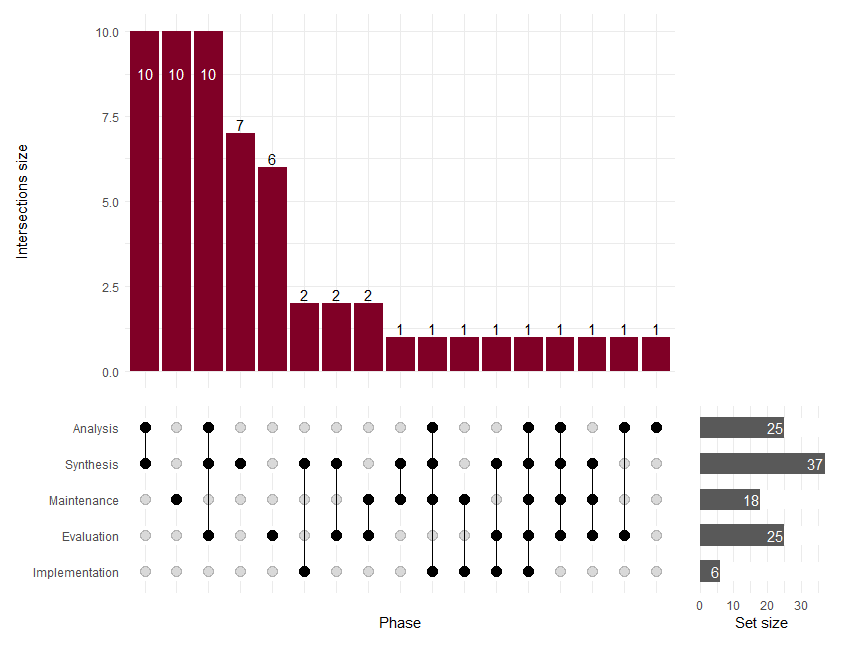}
	\caption{Architecting phases addressed in the primary studies and their intersections.}
	\label{fig:phases}
\end{figure*}
Further, we use the extracted keywords to identify the most popular topics covered by the studies in the field. 
Since the keywords themselves are defined at different levels of abstraction and exhibit a high degree of variability in the use of terms for the same concepts, we first organize them into categories.
Fig.~\ref{fig:keywords} presents the top ten such categories ordered by popularity.
The topics that are most frequently discussed fall within the \emph{architectural concepts} category, aggregating such topics as architectural principles, patterns, and tactics, design decisions and the respective principles and patterns, and software product lines and variability-related keywords.

Going beyond these simpler descriptive statistics, we next look into how to analyze and synthesize the remaining extracted data into a richer mapping of the state of the art.
The ``traditional'' way of achieving this result is through a bubble plot a la Petersen et al.~\cite{Petersen2008systematic}, with e.g.~the research facets, architecting phases, and sustainability dimensions as the axes of the plot, and the size of the bubbles signifying the number of studies in the intersection of the axes.
However, we have two categorical variables that can take multiple values at the same time (phases and dimensions), and one variable (contribution viewpoint) that is by definition correlated with another variable (sustainable software being usually an expression of technical sustainability~\cite{Venters2018}). 
As such, we feel that the bubble plot only diffuses the presented information instead of illustrating it, and we thus opt to present this information distributed across multiple complementary plots.

For the first part we are looking at the research facet of the primary studies, and their contribution viewpoint.
Fig.~\ref{fig:facets} synthesizes the extracted information from these fields to allow us to discuss the type of research conducted in the field.
From the figure we can conclude the following:
\begin{enumerate}
	\item While the majority of studies focus on sustainable software ($\sim64\%$), the ones aiming to deliver sustainability through software is not insignificant ($\sim36\%$). Given the fact that the former viewpoint is heavily associated with the technical dimension, this is somewhat expected.
	\item Sustainable software-focused studies dominate only two out of the three ``applied'' research facets, i.e.~\emph{solution proposals} and \emph{validation research}, with \emph{evaluation research} studies being split equally between the two viewpoints.
	This can be interpreted as a sign of both viewpoints perceived as the potential target for adoption by practitioners.
	\item There is a distinct lack of the more reflection-aiming types of studies (\emph{opinion papers}, \emph{experience papers}, and particularly \emph{philosophical papers}). This lack needs to be addressed: having more such studies in the future can facilitate further the maturation of the research topic. 
\end{enumerate}
Before moving on to also combine phases and dimensions, we first examine each of these extracted fields in isolation as the means of answering~\ref{rq:phases} and~\ref{rq:dims}, respectively.

\subsection{\ref{rq:phases} --- Architecting phases coverage}
\label{sec:phases}

%
\begin{figure*}
	\centering
	\includegraphics[width=.7\textwidth]{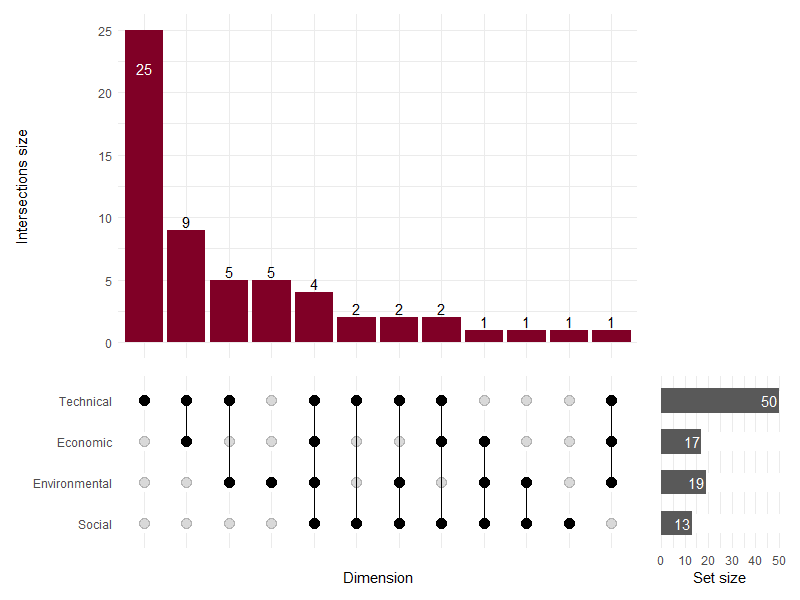}
	\caption{Sustainability dimensions addressed in the primary studies and their intersections}
	\label{fig:dims}
\end{figure*}
In order to answer~\ref{rq:phases} we summarize the architecting phases addressed by the primary studies and their intersections using an UpSet plot~\cite{lex2014upset}.
As seen in Fig.~\ref{fig:phases}, activities related to architecture synthesis are overall addressed the most, appearing in $\sim64\%$ of the studies.
However, maintenance (and evolution) is the most popular phase discussed \emph{individually}, as hinted already by the popularity of the respective keywords.
On the other hand, implementation, referring to the low level design of the system, is clearly the least-studied phase, discussed in only $\sim10\%$ of the studies. 
Also, we noticed that only 15 studies ($\sim26\%$) are addressing \emph{more than 3 phases} at the same time, and 10 out of these 15 are due to the prevalence of combining analysis-synthesis-evaluation. 
Furthermore, the majority of primary studies (37 out of 58, $\sim64\%$) are staying within the Hofmeister et al. model of architecting activities (i.e. analysis, synthesis, and evaluation, or a combination thereof).
Only 3 studies (a meager $5\%$) address 4 phases or more, and only 1 addresses all five.

\subsection{\ref{rq:dims} --- Sustainability dimensions coverage}
\label{sec:dims}

Similarly to the architecting phases, we use the UpSet plot of Fig.~\ref{fig:dims} to answer~\ref{rq:dims}.
As shown in the figure, the technical dimension is by far the most popular one in the selected primary studies, appearing in a staggering $\sim86\%$ of all studies. 
As a matter of fact, it is more popular than all other dimensions put together including their combinations, and it is also discussed in most combinations with any other dimension. 
Social, in turn, is the most under-addressed dimension (13 studies, $\sim22\%$), and almost always discussed in combination with other dimensions.
Similarly, the economic dimension, while discussed as much as the environmental one, is always treated together with another dimension and usually the technical one.
Perhaps more importantly, however, it appears that the focus on the technical dimension outweighs the number of approaches dealing with multiple dimensions: only 10 studies ($\sim17\%$) address 3 or more dimensions, and only 4 studies ($\sim7\%$) all four of them.

\subsection{Combining phases and dimensions}

\begin{figure}
    \centering
    \includegraphics[width=\columnwidth]{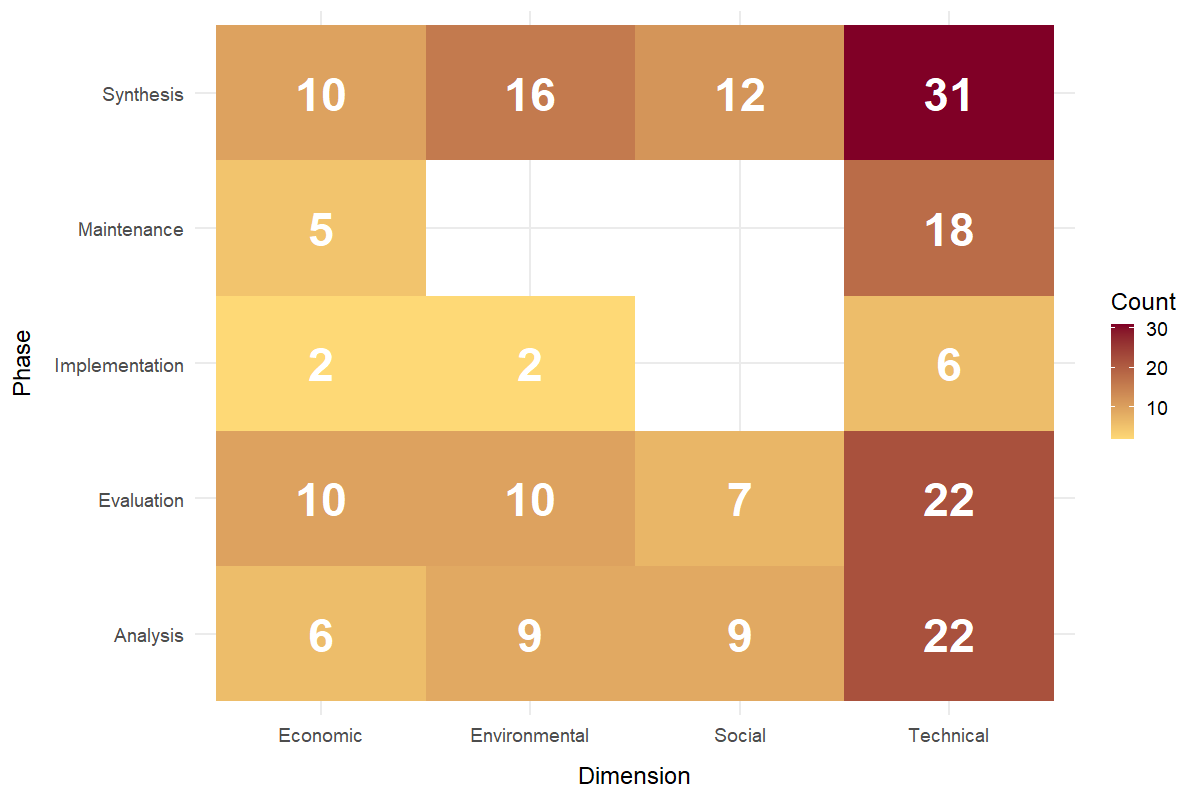}
    \caption{Co-occurrences of sustainability dimensions and architecting phases addressed by the primary studies}
    \label{fig:heatmap}
\end{figure}

Returning now to~\ref{rq:overall}, we plot the co-occurrences of the phases and dimensions appearing in each primary study as a heatmap in Fig.~\ref{fig:heatmap}.
This provides us with an eagle eye's view of their interactions. 
As shown in the figure, for example, \emph{only two dimensions are discussed across all phases of software architecting}: the \emph{technical and economic} ones; 
however, the former is much more popular than the latter 
(50 out of 58 studies, as discussed above). 
There are also obvious gaps in phases discussed in conjunction with the environmental and social dimensions, and a need for more research addressing the non-technical dimensions in general. 

\section{Discussion}
\label{sec:discussion}

In the previous section we presented our analysis of the state of the art of the intersection of sustainability and software architecture.
We stayed clearly away from identifying this area as ``sustainable software architecture'' since it has been historically associated with a very specific inward-facing meaning of sustainability in the literature associated with its resilience under change.
We focused on mapping out the most popular aspects of the state of the art, and more importantly, on identifying gaps. 
One point that was perhaps lost in this analysis is the set of possibilities created by the accumulated work in the area.
To be more specific: the number of available works and their apparent increasing maturity present an opportunity for the establishment of \emph{sustainability-aware architecture} as a separate field of study. 
Such field would concern itself first and foremost with the interaction between architecting activities from across the system lifecycle and the various dimensions of sustainability as a software-system quality property.

An important step towards realizing this vision is addressing first the diagnosed gaps in our knowledge of the topic.
To this aim, we identified a list of research items we find promising for the research community to work on in the coming years:
\begin{enumerate}[label=\textbf{Item \#\arabic*:},ref=Item \textsubscript{\arabic*},leftmargin=*,itemindent=3.5em,itemsep=0.5em]
    \item \emph{Actively pursuing reflection-oriented studies.} As discussed in Section~\ref{sec:rq1}, the literature seems to favor the more practical types of research at the expense of the more reflection-oriented ones such as opinion, experience, and philosophical papers.
    This difference should become smaller as the field continues to mature and more knowledge is accumulated by the involved researchers.
    However, actively pursuing more of these types of studies already offers the opportunity for performing a sanity check to a still growing field such as the one discussed here. 
    \item \emph{Wider coverage of the architecting lifecycle.} Figures~\ref{fig:phases} and~\ref{fig:heatmap} paint a picture of inequalities in the current works with respect to their treatment of phases.
    With the exception of works exclusively focusing on maintenance, there are few works going beyond the core Hofmeister et al. analysis-synthesis-evaluation combination of phases.
    More works addressing ideally the whole lifecycle are therefore necessary, or at least considering the implementation and maintenance phases, and specifically going beyond the intersection of the latter with the technical dimension.
    \item \emph{Going beyond the technical dimension.} In a similar manner, and by combining Fig.~\ref{fig:dims} with~\ref{fig:heatmap}, we can easily observe how pervasive is 
    the technical dimension in the current discourse.
    To some extent, this is to be expected: the notion of sustainable software itself has been equated for many publications with that of maintainability, adaptability, resilience, and other concepts under the technical sustainability umbrella~\cite{Venters2018}.
    However, and especially in the more recent years, the research community has been stressing the need for a multi-dimensional approach to dealing with sustainability as formulated by the Karlskrona manifesto~\cite{Becker2014}.
    As such, there is an urgent need for approaches incorporating as many dimensions as possible, and addressing especially the least popular ones (environmental, economic, and social).
    \item \emph{Towards a sustainability-aware architecting framework.} There is no primary study in our sample which covers all dimensions and phases at the same time. 
    This might have been understandable and acceptable while the field was still in its inception, but with its increasing maturity it becomes important to be able to deal with the multi-dimensionality of sustainability in a structured and holistic manner.
    Assuming that the identified gaps in the state of the art begin to be covered by the previous items in this agenda, an appropriate \emph{architecting framework} as discussed in~\cite{bookchapter} would then be necessary for integrating them into one coherent solution. 
    The most important element of this framework would be providing a connection between the various architecting activities and their effect, both direct and indirect, to each dimension.
    Outlining the vision for this framework is an important work item on its own. We plan to pursue this by extending the SAF Toolkit~\cite{SAFToolkit}.
\end{enumerate}

\section{Threads to validity}
\label{sec:validity}

For the discussion on the threats to the validity of this work, and the mitigating measures we took, we use Ampatzoglou et al.~\cite{ampatzoglou2019} as our guide.
Accordingly, we acknowledge potential issues along all three of the there-defined threat categories:

\noindent\emph{Study Selection Validity: }
This category aggregates threats rising from the first two phases of secondary studies, i.e.~search and filtering.
The publications used for the mapping process in this study were retrieved by querying 5 online databases using as broad query terms as possible.
The query string used was constructed in alignment with the PI(COC) search strategy, and was piloted over multiple iterations before converging to the form used in this study.
Duplicate studies were automatically removed from the final sample, and snowballing was performed to provide additional depth to our search.
In addition, selection criteria were defined following the research questions, and their application was performed first independently and then after consensus' building to minimize the bias.

\noindent\emph{Data Validity: }
This category includes threats applying to the data extraction and analysis phases of the secondary study. 
These threats are roughly organized in three groups: limitations of the dataset, data extraction bias, and research-introduced bias.
With respect to the first, and by including primary studies that propose (reusable) architectures in addition to works that study explicitly architecture as a topic, we increased both the size and heterogeneity of the dataset. 
To mitigate the issues that arise from data extraction, and similarly to the filtering process, we performed the extraction first independently and then developed a consensus through discussion among the researchers.
The senior researchers involved in this study have provided separate quality assessment to this goal.
Finally, to minimize research-introduced bias we adopted widely accepted conceptual frameworks (architecting phases defined by Tang et al.~\cite{tang2010comparative} and sustainability dimensions by Lago et al.~\cite{Lago2015}) for classification purposes.

\noindent\emph{Research Validity: }
This last category is concerned with the overall research design; it applies throughout the phases of the study and focuses mainly on two aspects: generalizability and repeatability.
With respect to the former, our research questions are motivated by both the lack of a similar review in the literature, and the necessity of the topic itself.
With respect to the latter, the replication package for the study has been made available online, even though at this stage as a shared private link only.

\section{Conclusion}
\label{sec:conclusion}

Software engineering studies have been increasingly concerned with the topic of sustainability, both as it is delivered by the systems they study, and of the software systems themselves.
A number of secondary studies from the last years are attesting to the existence of a vibrant community working on various related research topics.
One such topic, that of the intersection of software architecture and sustainability, is the focus of this work.
Having no other secondary study specifically addressing this topic, in this work we systematically map existing approaches from the literature on both sustainable architectures and sustainability through architecture.

Our findings show a quickly maturing research community that however needs to address specific deficiencies in its coverage of both architecting activities and sustainability dimensions.
Based on our analysis, we propose a list of research items for the following years towards the establishment of sustainability-aware (software) architecture as a new field of study.
The items in this research agenda are the focus of our future work.

\bibliographystyle{IEEEtran}
\bibliography{bibliography}

\end{document}